\documentclass[aps,prl,amsmath,amssymb,twocolumn]{revtex4-2}

\usepackage{graphicx}
\usepackage{adjustbox}
\usepackage{bm}
\usepackage{color}
\usepackage{braket}
\usepackage{standalone}
\usepackage{multirow}
\usepackage{tikz}
\usepackage{mathrsfs}
\usepackage[utf8]{inputenc}
\usepackage{dsfont}
\usepackage[colorlinks,bookmarks=true,citecolor=blue,linkcolor=red,urlcolor=blue]{hyperref}

\usepackage{bbm}
\usepackage[caption = false, labelformat=simple]{subfig}
\usepackage{floatrow}
\usepackage{verbatim}

\begin{document}

\title{Non-Linear Interference Challenging Topological Protection of Chiral Edge States}
\author{Benjamin Michen}
\email{benjamin.michen@tu-dresden.de}
\author{Jan Carl Budich}
\email{jan.budich@tu-dresden.de}
\affiliation{Institute of Theoretical Physics${\rm ,}$ Technische Universit\"{a}t Dresden and W\"{u}rzburg-Dresden Cluster of Excellence ct.qmat${\rm ,}$ 01062 Dresden${\rm ,}$ Germany}
\date{\today}

\begin{abstract}
We report on a non-linear scattering effect that challenges the notion of topological protection for wave packets propagating in chiral edge modes. Specifically, in a Floquet topological system close to resonant driving and with a non-linear potential, we demonstrate how a wave packet propagating in a chiral edge mode may be irreversibly deflected by scattering off a localized wave-packet, or pass the collision region virtually unaffected in an approximately linear fashion. An experimentally accessible knob to tune between those two scenarios is provided by the relative phase between the involved wave-packets. This genuinely non-linear interference phenomenon is in stark contrast to linear scattering off a static impurity, which cannot destroy a topological edge state. Besides corroborating our findings with numerically exact simulations, we propose two physical platforms where our predictions may be verified with state of the art experimental techniques: First, a coupled waveguide setting where non-linearity has been engineered via an intensity-dependent optical index. Second, a Bose-Einstein condensate of cold atoms in an optical Honeycomb lattice governed by a non-linear Gross-Pitaevskii equation that effectively accounts for many-body interactions.
\end{abstract}

\maketitle

{\it Introduction.---} 
As a direct consequence of linear equations of motion, the superposition principle is ubiquitous in physics, with ramifications ranging from classical waves to quantum wavefunctions. Non-linear terms violating the superposition principle are often introduced to model complex interactions at a simpler effective level \cite{NL_acoustics, NL_optics, GP_1, GP_2}. There, intriguing phenomena and applications arise, such as  quantized vortices in Bose-Einstein condensates (BECs) \cite{ GP_3, GP_4}, self-focussing of laser beams \cite{Self_focussing_1, Self_focussing_2}, or novel diagnostic procedures \cite{NL_diagnostic_1, NL_diagnostic_2}.

Very recently, non-linear effects have been identified in the realm of topological matter  \cite{NL_phase_transition, NL_attractor_repeller, NL_instability, NL_edge_states, NL_edge_states_2, Landau_photon_polaritons, NL_topological_laser, Topo_frequency_comb}, even though the underlying topological classification schemes are based on linear systems \cite{HasanKane2010,Qi2011,Wen2017,Armitage2018}. Specific manifestations have been found to occur in a variety of settings in photonics, \cite{NL_photonics_1, NL_photonics_2, NL_photonics_3, NL_photonics_4, NL_photonics_5}, mechanics \cite{NL_mechanics_1, NL_mechanics_2, NL_mechanics_3}, and electrical circuits \cite{NL_electronics_1, NL_electronics_2}. Non-linear phenomena discovered in this context include self-localization \cite{NL_self_localization} and edge solitons \cite{NL_solitons_1, NL_solitons_2, NL_solitons_3}. Furthermore, optical non-linearities have been observed to be capable of affecting photonic band structure parameters so as to induce intensity-dependent linearized topological properties \cite{NL_induced_topology}. 

\begin{figure}[htp!]	 
{\includegraphics[trim={0.75cm 0cm 0.75cm 0.cm}, width=0.95\linewidth]{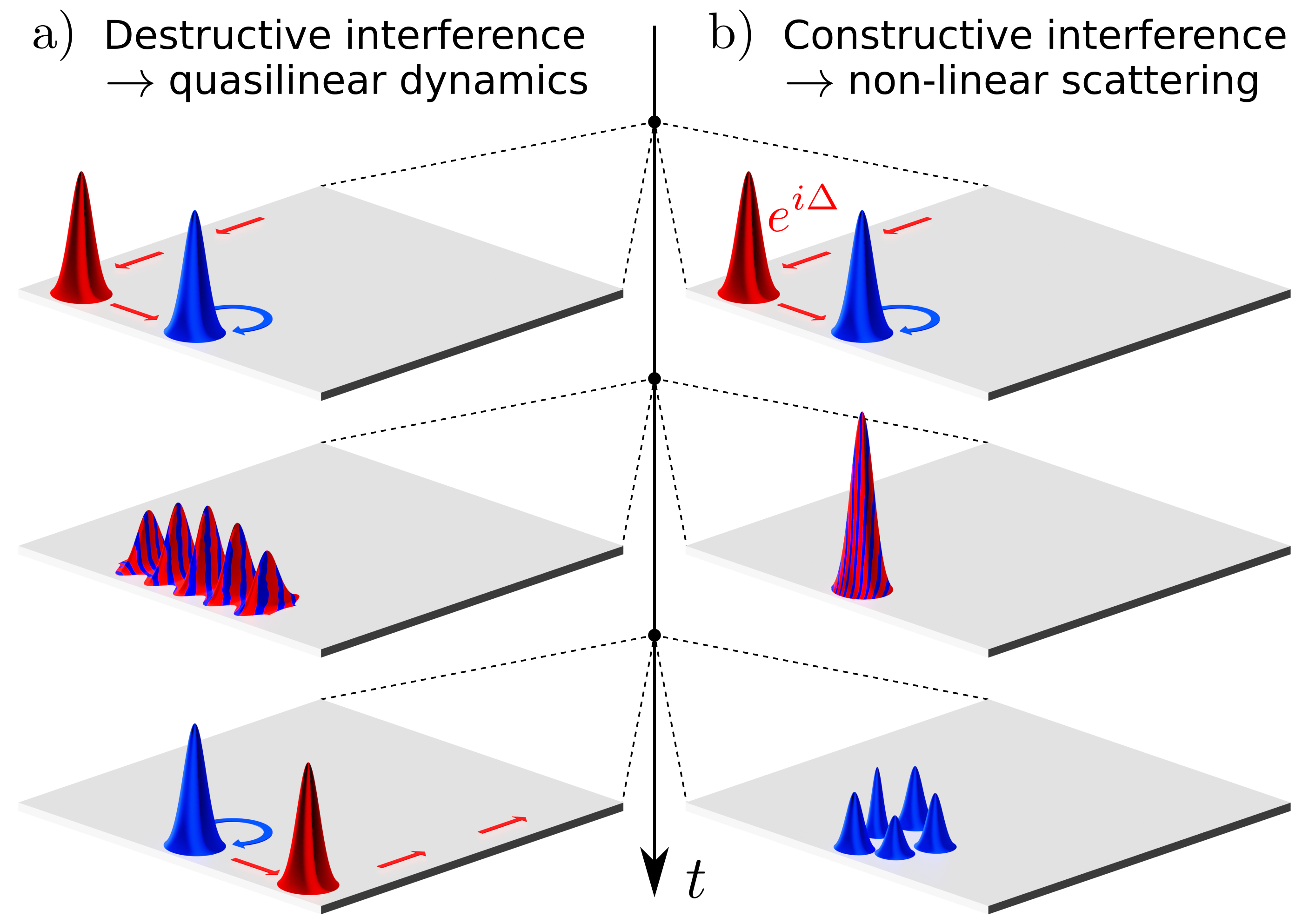}}
\caption{Illustration of a non-linear topological Floquet system (cf.~Eq.~(\ref{Eqn:NL_SGL})), where an edge mode (red) may propagate uninhibited up to a certain intensity threshold. If it passes a locally circulating wave packet (blue), the dynamics heavily depends on the interference between the wave packets. (a) Destructive interference keeps the overall intensity below the threshold so as to maintain quasi-linear dynamics within the realm of edge protection by linear Floquet topology. (b) Constructive interference escalates the peak intensity above the threshold so as to cause strongly non-linear scattering in violation of both the superposition principle and topological protection. The initial relative phase $e^{i \Delta}$ provides a knob to switch between scenarios (a) and (b). 
 }\label{Fig:illustration}
\end{figure}

Here, we reveal  a striking violation of the superposition principle characterizing the non-linear dynamics of wave packets colliding in a Floquet chiral edge mode. This intriguing interplay of topology and genuine non-linearity depends crucially on the interference between the involved wave-packets which determines the overall peak intensity in the collision region. Concretely, by tuning the initial relative phase of two wave packets, qualitatively distinct scenarios may be observed (see Fig.~\ref{Fig:illustration} for an illustration): for destructive interference in the collision region, linear dynamics hallmarked by the superposition principle prevails (see Fig.~\ref{Fig:illustration}(a)). By contrast, in the case of constructive interference, the chiral edge state perishes above a threshold peak intensity, thus overwriting the topological protection of the incident wave packet irreversibly (see Fig.~\ref{Fig:illustration}(b)).

As a platform, we consider a Floquet system with a non-linear on-site potential that scales with the density of the wave function. The linear part of the system is tuned to resonant driving \cite{res_driving_1, res_driving_2}, which allows for a perpetually circulating bulk excitation. Within an anomalous Floquet topological phase \cite{Floquet_Majoranas, Floquet_Topology, Anomalous_BBC}, non-linear scattering may then occur between a wave-packet in the chiral edge state and the aformentioned circulating bulk mode (cf.~Fig.~\ref{Fig:illustration}). Floquet systems with anomalous topology and intrinsic non-linearity have been realized with state-of-the-art experimental technology in various physical platforms \cite{NL_induced_topology, Photonic_quantum_walk, Anomalous_Photonics, Anomalous_Honeycomb}. In the following, after a general theoretical analysis, we will illustrate our findings with two model systems of immediate experimental relevance, a square lattice model of coupled waveguides \cite{review_photonics_1,  review_photonics_2, Midgap_states, NL_induced_topology}, and a honeycomb model for ultracold bosonic atoms where the dynamics of BEC wave-packets is highly controllable \cite{Anomalous_Honeycomb}.

{\it Non-linear Floquet dynamics of wavepackets.---}
We consider a system governed by the non-linear generalized Schrödinger equation 

\begin{align}
i \frac{\mathrm d}{\mathrm d t} \psi_j(t) = \sum_{j'} H_F(j,j',t) \psi_{j'}(t) + \gamma |\psi_j(t)|^2 \psi_j(t), \label{Eqn:NL_SGL}
\end{align}
where $\psi_j(t)$ are the spatial components of the state vector on a lattice and units are chosen such that $\hbar = 1$. Here, $H_F(t)$ denotes a linear Floquet Hamiltonian with period $T$, and the non-linear term with coupling strength $\gamma$ may be seen as an on-site potential that scales with the density $|\psi_j|^2 $. A non-linearity of this type emerges naturally in a variety of physical platforms, for instance through an intensity-dependent optical index in photonic systems \cite{NL_photonics_5} or from a mean-field treatment of weak interactions in a BEC, which yields the Gross-Pitaevskii equation \cite{GP_1, GP_2, GP_3, GP_4}.

We start by discussing on a general note the non-linear interference mechanism in the dynamics of two wave packets  $\Psi$ and $\Phi$ that is at the heart of our present analysis. Let $\Psi_{ j}(t)$ and $\Phi_{ j}(t)$ be two solutions to Eq.~(\ref{Eqn:NL_SGL}). Then, $\mathrm{e}^{i \Delta} \Phi_{ j}(t)$ is also a solution for any $\Delta \in \mathbb R$, but the superposition $g_{ j}(t) = \Psi_{ j}(t) + \mathrm{e}^{i \Delta} \Phi_{ j}(t)$ generally has a different intensity distribution and thus does not satisfy Eq.~(\ref{Eqn:NL_SGL}). In particular, the local intensity of  $g_{ j}(t)$ is highly sensitive to the relative phase $e^{i \Delta}$, which can be exploited to tune the importance of the non-linearity in the combined dynamics of $\Psi$ and $\Phi$. Specifically, we find for the time-dependence of the superposition 

\begin{align}
i \frac{\mathrm d}{\mathrm d t} g_{ j}=& \sum_{j'} H_F(j,j',t) g_{j'} + \gamma |g_{ j}|^2 g_{ j} \nonumber \\
& - \gamma \left( 2 \mathrm{Re}\left [ \Psi^*_{j} \mathrm{e}^{i \Delta} \Phi_{ j} \right] g_{ j}
 + \Psi_{j}  \mathrm{e}^{i \Delta} \Phi_{ j} g_j^* \right ),  \label{Eqn:dg_dt}
\end{align}
where the time argument has been suppressed for brevity, and the excess terms as compared to Eq.~(\ref{Eqn:NL_SGL}) in the second line describe the non-linearity induced violation of the superposition principle. While the superposition principle is always fulfilled as long as the two wave-packets do not overlap, the second line of Eq.~(\ref{Eqn:dg_dt}) can be of crucial importance in the collision region of $\Psi$ and $\Phi$. Assuming the initial $\Psi_{j}$ and $\Phi_{ j}$ as Gaussian wavepackets with a homogeneous phase around their peaks, the relative phase $\Delta$ can be used to minimize or maximize the peak intensity of $g_j(t) = \Psi_{j}(t)  + \mathrm{e}^{i \Delta} \Phi_{ j}(t)$ during the time of the overlap. Thus, depending on the choice of $\Delta$ the superposition principle may remain to good approximation in place  (cf. Fig.~\ref{Fig:illustration}a) or may be drastically violated by maximizing the peak amplitude, thus leading to strong interaction between the two wavepackets (cf. Fig.~\ref{Fig:illustration}b).

Now, we turn to discussing the intriguing interplay of non-linearity and chiral edge states, i.e. we assume $H_F(t)$ to be a topologically non-trivial Floquet Hamiltonian \cite{Anomalous_BBC}. There, the total system described by Eq.~(\ref{Eqn:NL_SGL}) will generally support chiral edge states only for intensities below a certain threshold. For higher amplitudes of the wave function, the non-linear term will effectively act as a dynamical sublattice potential that breaks the topological phase. A complementary type of intensity dependent Floquet topology has recently been experimentally observed in an array of optical waveguides \cite{NL_induced_topology}, where edge transport is enabled by higher light intensities.

Here, inspired by Ref.~\cite{NL_induced_topology}, we analyze the dynamical interaction of wave-packets so as to predict the irreversible breakdown of topological protection due to the aforementioned non-linear interference mechanism. To this end, we consider as initial condition a wave packet $\Psi$ propagating in a chiral edge state and a pulse $\Phi$ circulating close to the edge (see Fig.~\ref{Fig:illustration}). While the existence of chiral edge states is guaranteed by the topology of $H_F$, the stationarity of the circulating bulk mode relies on a resonant driving condition \cite{res_driving_1, res_driving_2}. The individual intensities of both pulses are below the intensity threshold that would break the topological phase. Depending on the total relative phase $\Delta$ of the two wave-packets, we predict the edge mode to either pass the circulating pulse virtually unaffected or to be strongly scattered into localized bulk states, as will be corroborated with numerical simulations on two microscopic lattice models below. 

{\it Square lattice model---} 
We illustrate our findings on the basis of a non-linear version of a minimal square lattice model for an anomalous Floquet insulator proposed by Rudner et al.~\cite{Anomalous_BBC}. This model has been realized in recent experiments using coupled waveguide arrays, also in its non-linear form \cite{Anomalous_Photonics, NL_induced_topology}. There, the driving period is divided into four steps of equal length, and the hoppings $c_j(t)$, $j = 1,2,3,4$ along the different nearest-neighbor (NN) links are set to $c$ during the $j$-th step, respectively, and zero otherwise. The linear part $H_\mathrm F ^\mathrm{phot}(t)$ of the model is thus entirely determined by the driving period $T$, the lattice spacing $a$, and the hopping amplitude  $c$. The system has two sublattices $A$, $B$, and its bulk is described in reciprocal space by the Bloch Hamiltonian 

\begin{align}
H_\mathrm F ^\mathrm{phot}(\bm k, t) = \sum_{j = 1} ^4  
\begin{pmatrix}
 0 & c_j(t) \mathrm e^{i \bm b_j \bm k} \\
 c_j(t) \mathrm e^{-i \bm b_j \bm k} & 0
 \end{pmatrix}
 \label{Eqn:H_F_Phot}
\end{align}
with $\bm b_1 = - \bm b_3 = (a, 0)$ and $\bm b_2 = - \bm b_4 = (0, a)$, for more details see \cite{SOM}. In the following, we use units where $a = 1$.

\begin{figure}[htp!]	 
\includegraphics[ trim={0.cm 0cm 0cm 0.cm}, width=0.95\linewidth]{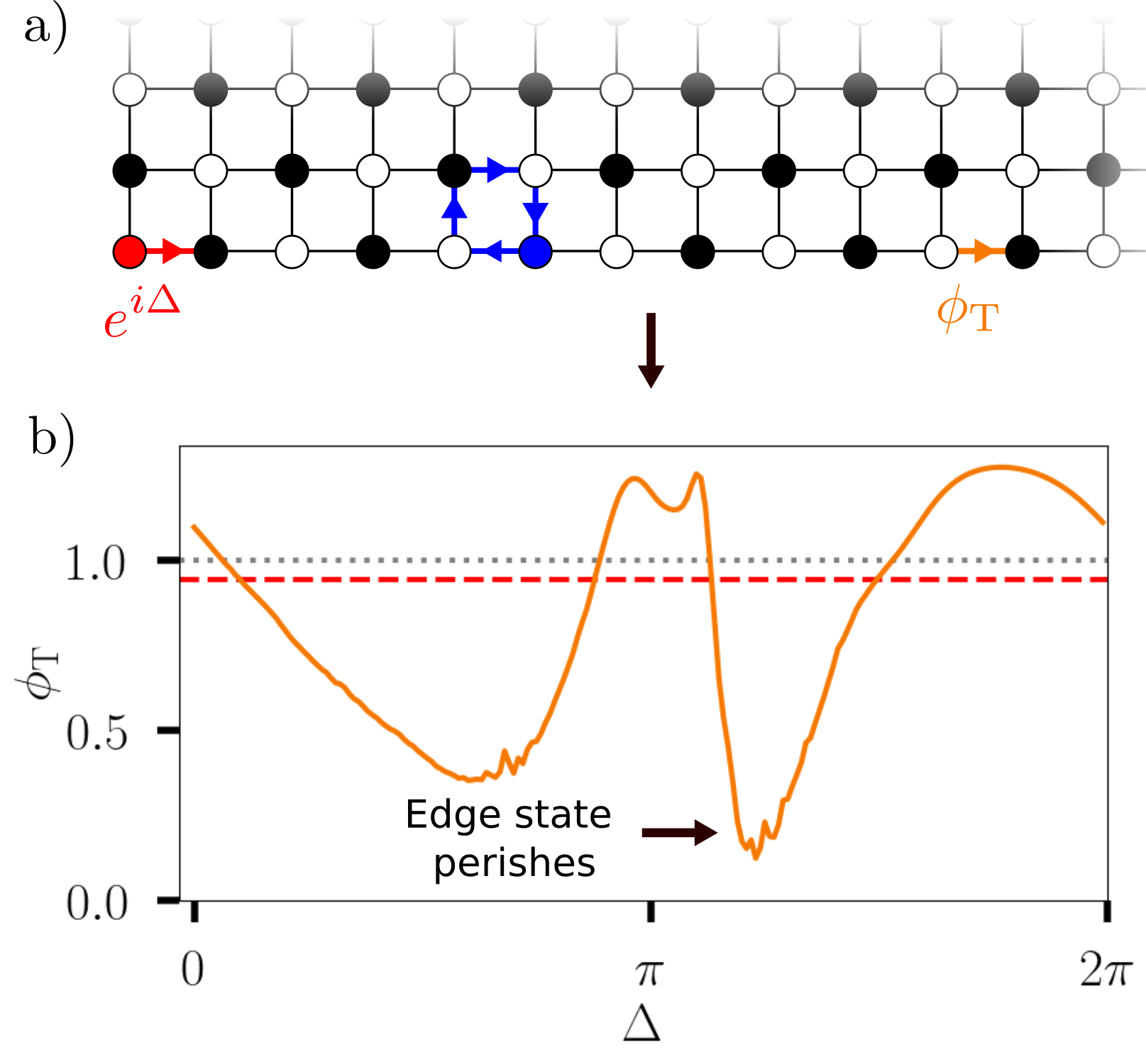} 
\caption{(a) Schematic of square lattice model (see Eq.~\ref{Eqn:H_F_Phot}) with sublattices $A$ and $B$ indicated by white and black circles, respectively. System size is $16 \times 16$ sites and parameters are $c = \frac{2 \pi}{T}, \gamma = \frac{3}{T}$. Initial excitations as per Eq.~(\ref{Eqn:Init_cond_phot}) are located on the lower left edge, with chiral edge state amplitude $\mathrm  e^{i \Delta}$ (red) and the circulating bulk state amplitude $1$ (blue). We measure the transmitted probability flux $\phi_\mathrm{T} = \int_0^{10T} \phi_{\bm j, \bm \delta}(t) \mathrm d t$ as per Eq.~(\ref{Eqn:prob_flux}) through the link marked in orange. (b) $\phi_\mathrm{T}$ as a function of the relative phase $\Delta$ between the red and blue pulse. The flux generated by the edge state (red pulse) alone in the non-linear system ($\gamma = \frac{3}{T}$) is indicated as a red dashed line and in the linear system ($\gamma = 0$)  as a grey dotted line for reference.
 }\label{Fig:phot_model_flux}
\end{figure}

Resonant driving is achieved for $c = \frac{2 \pi}{T}$, which places the linear part of the system in an anomalous topological Floquet phase. There, the winding number $W$~\cite{Anomalous_BBC} of the bulk system (cf. Eq.~(\ref{Eqn:H_F_Phot})) is given by $W=1$, such that a single counterclockwise propagating edge mode appears in a finite geometry. We set the strength of the on-site nonlinearity to  $\gamma = \frac{3}{T}$ and solve the resulting non-linear Schrödinger Equation (cf. Eq.~(\ref{Eqn:NL_SGL})) for a finite lattice of $16 \times 16$ sites numerically. For this value of $\gamma$, a pulse of up to unit intensity can still propagate virtually uninhibited. Hence, as sketched in Fig.~\ref{Fig:phot_model_flux}a, our initial conditions consist of an edge state (red) with amplitude $\mathrm  e^{i \Delta}$ and a circulating bulk state (blue) with amplitude $1$.  Indexing the physical sites of the lattice by $j_x$, $j_y$ (irrespective of the $A$, $B$ sublattice structure), this can be written as

\begin{align}
\psi_{j_x, j_y}(t = 0) = \mathrm e^{i \Delta} \delta_{jx, 0}\delta_{jy, 0} + \delta_{jx, 0} \delta_{jy, 5}. \label{Eqn:Init_cond_phot}
\end{align}

For these initial conditions, our numerical calculations show that the dynamics are highly sensitive to the relative phase $\Delta$ of the pulses. To quantify the influence of $\Delta$ on the non-linear interaction between the wave packets, we define the flux density through a lattice link starting from a continuity equation. Consider a non-linear time evolution as per  Eq.~(\ref{Eqn:NL_SGL}) with some $H_F(t)$ containing only time-dependent NN hoppings $J_{\bm \delta}(t)$. The density then obeys

\begin{align}
\frac{\mathrm d}{\mathrm d t} |\psi_{\bm j}(t)|^2 = \sum_{\bm \delta \in \mathrm {NN}} \underbrace{2 \mathrm {Im} [J_{\bm \delta} (t) \psi^*_{\bm j}(t) \psi_{\bm j + \bm \delta}(t)]}_{-\phi_{\bm \delta, \bm j}(t)}, \label{Eqn:prob_flux}
\end{align}
where $\phi_{\bm \delta, \bm j}$ can be interpreted as the probability flux density leaving site $\bm j $ through the $NN$ link $\bm \delta$. This expression holds for any lattice type and a generalization to longer-ranged hoppings is straightforward. Now, we pick a link further down the edge, i.e. far away from the scattering region in propagation direction of the edge state. Specifically, we choose the link connecting the 11th and 12th site of the lower edge (cf.~Fig.~\ref{Fig:phot_model_flux}a), and integrate the flux over a sufficiently long time such that the entire remainder of the edge pulse has passed. Here, an integration time of ten driving periods is sufficient to capture all of the transmitted flux, i.e. $\phi_\mathrm{T} = \int_0^{10T} \phi_{\bm j, \bm \delta}(t) \mathrm d t$.

Fig.~\ref{Fig:phot_model_flux}b shows the resulting $\phi_\mathrm{T}$ as a function of $\Delta$, with the total flux generated by the wave packet in the edge mode alone indicated as a red dashed and grey dotted line for $\gamma = \frac{3}{T}$ and $\gamma = 0$, respectively, for comparison. The deviation of $\phi_\mathrm{T}$ from the red dashed line thus quantifies the violation of the superposition principle by non-linear scattering of the two wave-packets, while the deviation of the red dashed line from the grey dotted line is a correction to the ideal linear Floquet topology at finite intensity. Depending on the relative phase, we observe that the edge mode can either be scattered almost completely into the bulk (the dip around $\Delta \approx 1.2 \pi$ marked by a black arrow in Fig.~\ref{Fig:phot_model_flux}b, see also Fig.~\ref{Fig:illustration}b), pass the circulating bulk mode unaffected (the intersections with the red dashed line), or even scatter a portion of the circulating mode so as to amplify the edge state propagation (the peaks at $\Delta \approx \pi$ and $\Delta \approx 1.8 \pi$). 

{\it Honeycomb model---} 
To illustrate the generality of our findings, we also investigate a periodically driven optical Honeycomb lattice that hosts a BEC. This system has been realized in Ref.~\cite{Anomalous_Honeycomb} and can be tuned to exhibit a similar effect to the square lattice model considered before. The system is effectively described by a tight-binding Honeycomb lattice with two atoms per unit cell and time-dependent hoppings 

\begin{align}
J_l (t) = A  e^{B \cos(\omega t + (l-1) \phi)} + C, \; \phi  = \frac{2 \pi}{3}, \label{Eqn:hopping_HC}
\end{align}
along the directions $l=1,2,3$, where the parameters $A$, $B$, $C$ can be tuned externally, for more details refer to Ref.~\cite{SOM}. The driving period is $T = \frac{2 \pi}{\omega}$.

To ensure resonant driving and facilitate a stable locally circulating excitation, it is necessary that only one NN link is active at a time. Here, this can be achieved by picking a large value of $B = 7 $ in order to localize the exponentials in Eq.~(\ref{Eqn:hopping_HC}) and setting $C=0$. Furthermore, the ratio of the driving period and the overall hopping amplitude has to be chosen such that 100~\% intensity is transferred during link activation time, which amounts to taking $A = 0.0015 \frac{2 \pi}{T}$ \cite{SOM}. For these parameters, the linear part of the system enters an anomalous Floquet topological phase with $W=1$ protecting a counter-clockwise propagating edge state in a finite geometry. Turning on interactions leads to a non-linearity similar to Eq.~(\ref{Eqn:NL_SGL}) in the Gross-Pitaevskii equation governing the dynamics of the BEC \cite{GP_1, GP_2, GP_3, GP_4}, which is readily tunable through Feshbach resonances \cite{FB_resonance}. We set $\gamma = 1.5 \frac{2 \pi}{T}$, which allows for basically uninhibited propagation of pulses with up to unit intensity.

\begin{figure}[htp!]	 
\includegraphics[trim={0.cm 0cm 0.cm 0cm},clip,  width=0.95\linewidth]{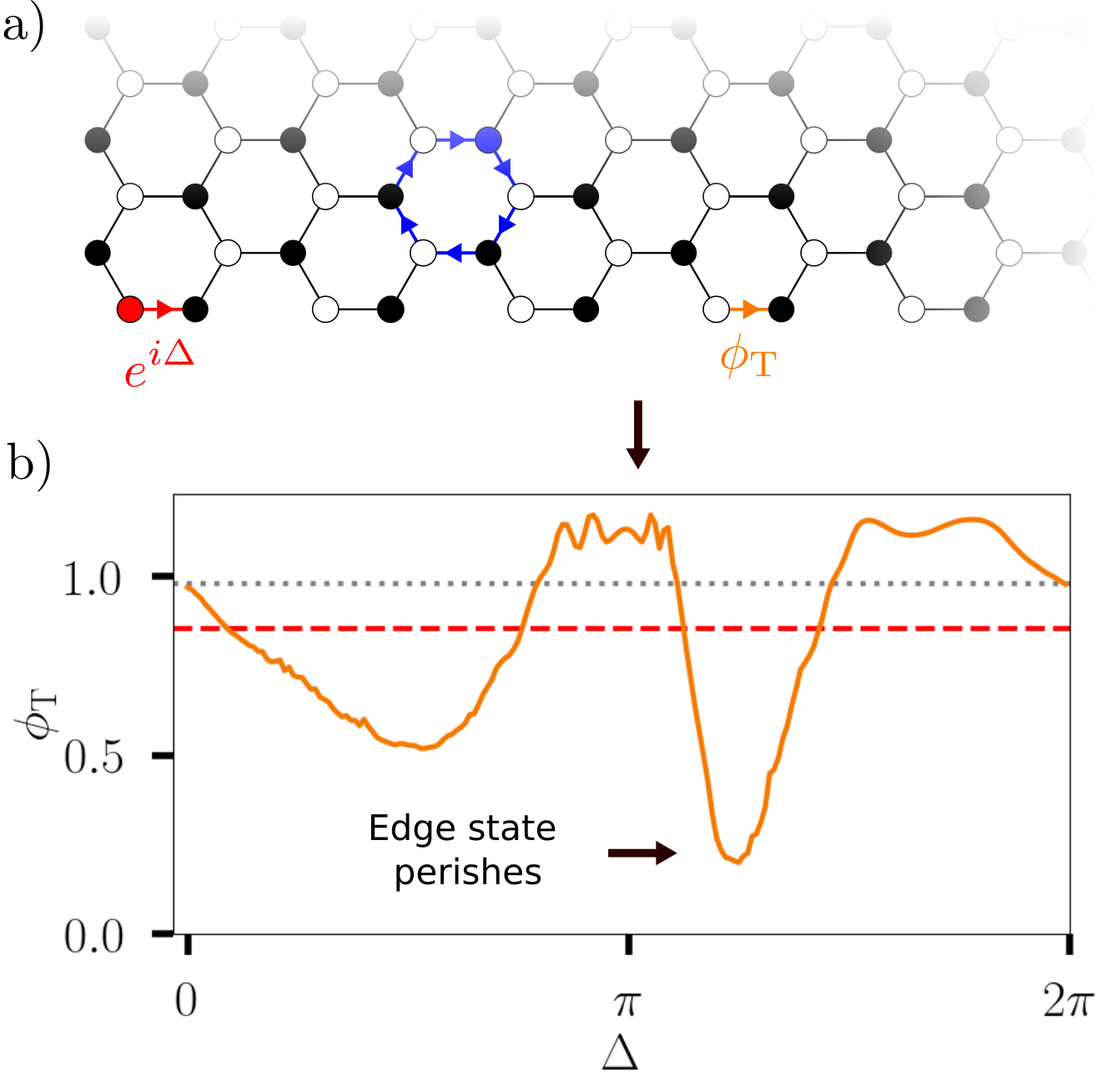} 
\caption{a) Schematic of Honeycomb model (see Eq.~\ref{Eqn:hopping_HC}). System size is $10 \times 10$ unit cells and further parameters are $A = 0.0015\frac{2 \pi}{T}$, $B = 7$, $C = 0$,  $\gamma = 1.5 \frac{2 \pi}{T}$. Initial excitations of the edge state amplitude $\mathrm  e^{i \Delta}$ (red) and the locally circulating state of amplitude $1$ (blue) as well as the transmitted probability flux $\phi_\mathrm{T} = \int_0^{20 T} \phi_{\bm j, \bm \delta}(t) \mathrm d t$ over 20 driving periods $T = \frac{2 \pi}{\omega}$ (cf.~Eq.~(\ref{Eqn:prob_flux})) through a link (orange) are indicated. (b) $\phi_\mathrm{T}$ as a function of the relative phase $\Delta$ between the red and blue pulse. The flux generated by the edge state (red pulse) alone in the non-linear system ($\gamma = 1.5 \frac{2 \pi}{T}$) is indicated as a red dashed line and in the linear system ($\gamma = 0$)  as a grey dotted line for reference.}\label{Fig:HC_model_flux}
\end{figure}

Following a similar protocol as before, we consider a lattice of $10 \times 10$ unit cells and start an edge state and a circlating bound state as illustrated in Fig.~\ref{Fig:HC_model_flux}a, with a relative phase $e^{i \Delta}$. The measurement of the transmitted probability flux over 20 driving periods  as per Eq.~(\ref{Eqn:prob_flux}) through a link down the edge (cf. Fig.~\ref{Fig:HC_model_flux}a) yields a largely similar result to the square lattice system which is shown in Fig.~\ref{Fig:HC_model_flux}b. The total flux generated by the wave packet in the edge mode alone is indicated again as a red dashed and grey dotted line for $\gamma = 1.5 \frac{2 \pi}{T}$ and $\gamma = 0$, respectively. The grey dotted  line deviates slightly from one because the system is not perfectly tuned to resonant driving \cite{SOM} such that a small portion of the initial unit intensity pulse escapes into the bulk. Analogous to the square lattice model of coupled waveguides, by tuning $\Delta$ we can mostly destroy the edge state ($\Delta \approx 1.2 \pi$, marked by black arrow in Fig.~\ref{Fig:HC_model_flux}b), let it pass relatively unaffected (the intersections with the red dashed line), or amplify the edge current (the peaks around $\Delta \approx \pi$ and $\Delta \approx 1.8 \pi$).

{\it Conclusion---} 
We have indentified and analyzed a non-linearity induced scattering mechanism that can be observed in topological Floquet systems with existing experimental techniques. In particular, a wavepacket propagating in a chiral edge state can scatter off another wavepacket, which can almost entirely destroy the edge state. By contrast to scattering off static impurities which merely locally redirect the edge state, the circulating wave packet should be seen as an effectively time-dependent impurity potential, for which protection of edge states is no longer guaranteed. Saliently, this effect can be controlled by an initial relative phase between the wavepackets, which provides an accessible knob to switch the non-linear scattering on or off. To illustrate the general applicability of our findings, we have presented qualitatively similar results on two microscopic lattice models representing recently implemented experimental settings in the context of coupled waveguide arrays and ultracold bosonic atoms, respectively.

We note that the collision of counter-propagating edge solitons in a non-linear Floquet system was investigated in an earlier work \cite{soliton_collision}. There, the non-linearity effectively modifies the dispersion of edge states from different band gaps such that solitons with opposite direction of propagation on the same edge emerge. However, this effect requires a specific intensity regime that is not able to break the topological phase even at constructive interference of the pulses, such that the solitons in this scenario are robust against scattering and varying their relative phase only induces a small spatial shift.

On a more general note, chiral edge states as one-way streets for transport and information processing may be seen as a promising platform for quantum- and nano-technology. To this toolbox, our present findings add a simple experimental switch for blocking and opening propagation through a chiral edge state by means of non-linear scattering. Within our present description in terms of non-linear wave-equations, non-linearity emerges at a mean-field level. In this light, identifying a counterpart to our findings in the quantum regime, where the interplay of many-body correlations and Floquet driving may qualitatively affect the properties of chiral edge modes, in our view defines an interesting direction of future research.   

\acknowledgments
{\it Acknowledgments.---}
We would like to thank Alexander Szameit and Tommaso Micallo for discussions. We acknowledge financial support from the German Research Foundation (DFG) through the Collaborative Research Centre SFB 1143, the Cluster of Excellence ct.qmat, and the DFG Project 419241108. Our numerical calculations were performed on resources at the TU Dresden Center for Information Services and High Performance Computing (ZIH).

\appendix

\onecolumngrid
\section{Supplementary Material for ``Non-Linear Interference Challenging Topological Protection of Chiral Edge States'' }
\twocolumngrid

\section{Details on the square lattice model} 

\begin{figure}[h!]	 
\includegraphics[ width=0.95\linewidth]{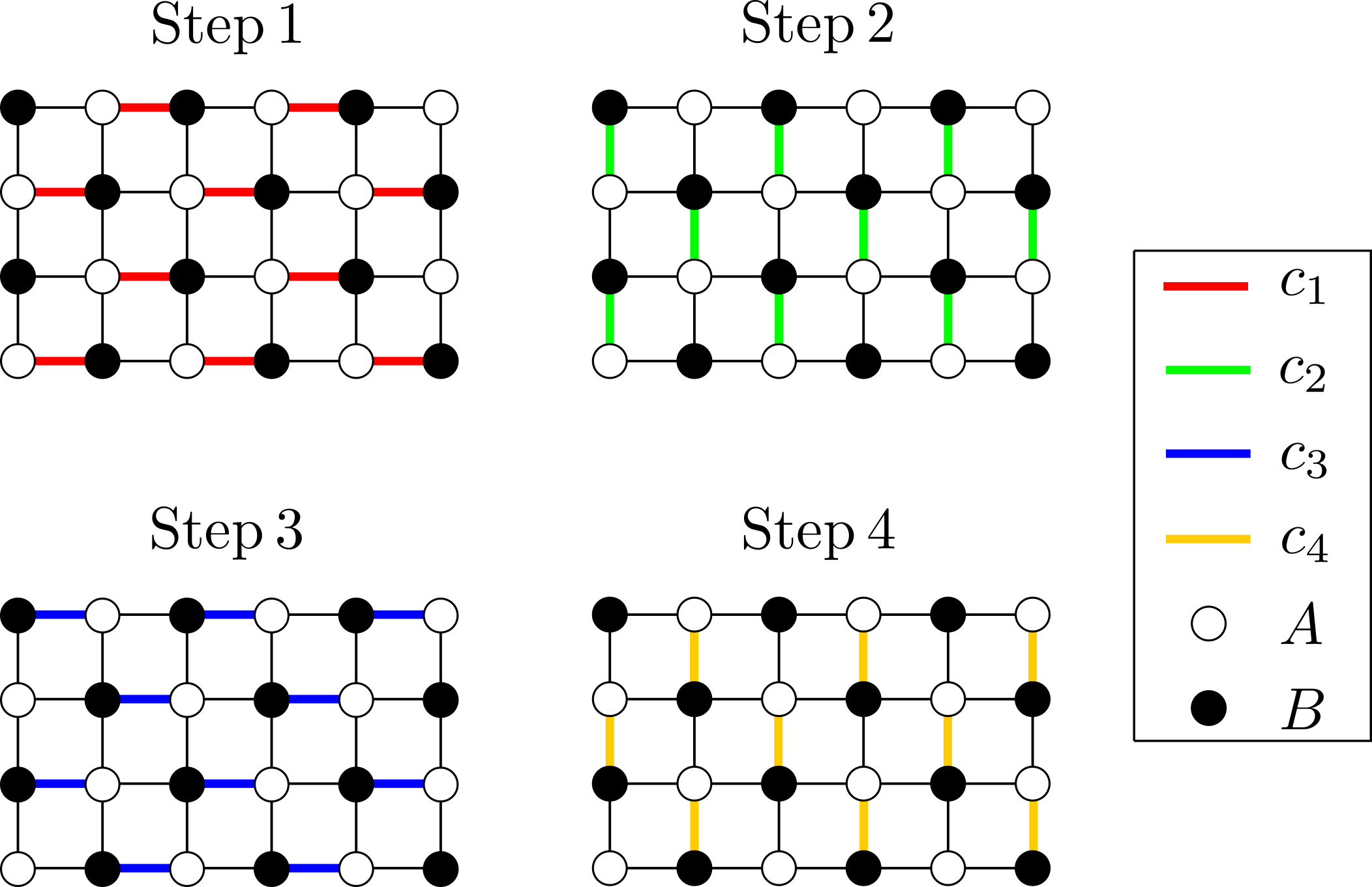} 
\caption{The linear Hamiltonian of the square lattice model consists of two sublattices $A$ and $B$ with time-dependent hoppings $c_j(t)$, $j = 1, 2,3,4$ along the different nearest-neighbor links. The spacing between the sites is $a$ and the driving period $T$ is devided into four steps of equal length. During the $j$-th step, the hopping $c_j(t)$ is set to $c$, otherwise to zero. The illustration shows a small slab of the model during each of the driving steps with the active link highlighted by color.
 }
\end{figure}

\section{Details on the Honeycomb model}
\begin{figure}[h!]	 
\includegraphics[ width=0.95\linewidth]{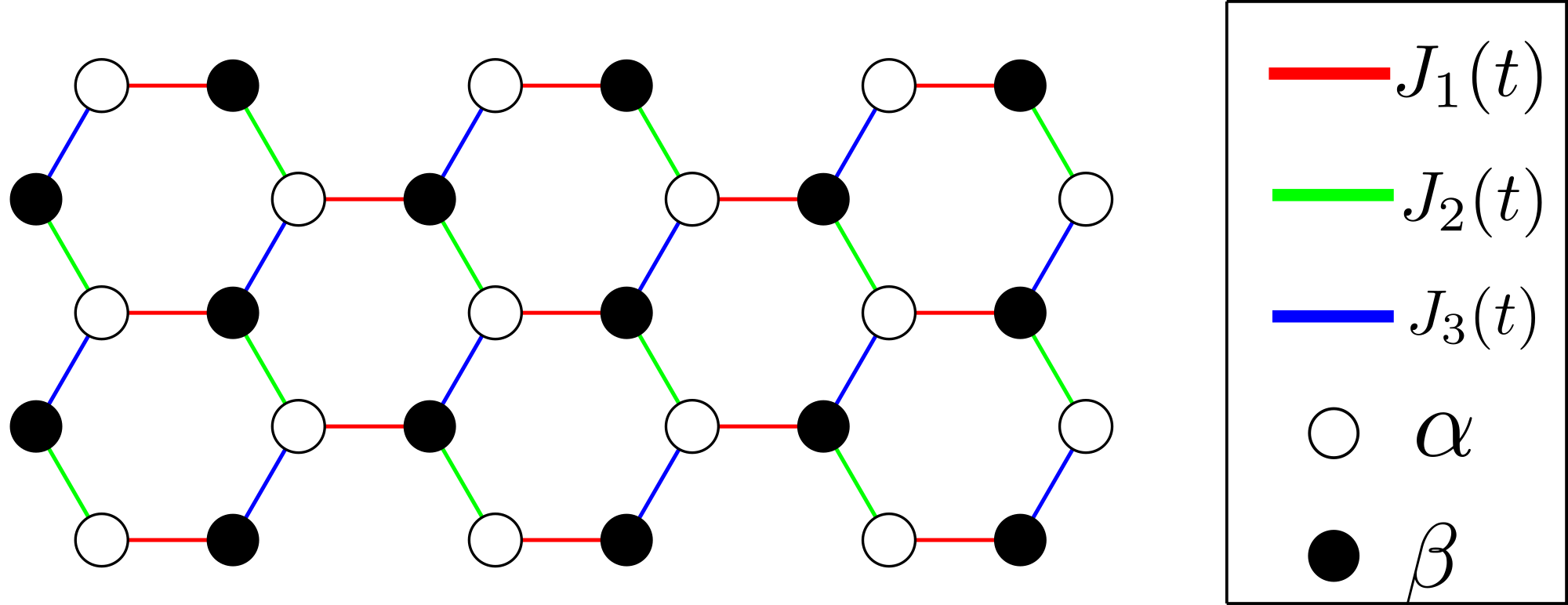} 
\caption{The linear Hamiltonian  of the Honeycomb lattice model consists of two sublattices $\alpha$ and $\beta$ with time-dependent hoppings $J_l(t)$, $l = 1, 2,3$ (see Eq.~\ref{APP:Eqn:hopping_HC}) along the different nearest-neighbor links.} \label{App:Fig:HC_model}
\end{figure}

\begin{figure}[h!]	 
\includegraphics[ width=0.95\linewidth]{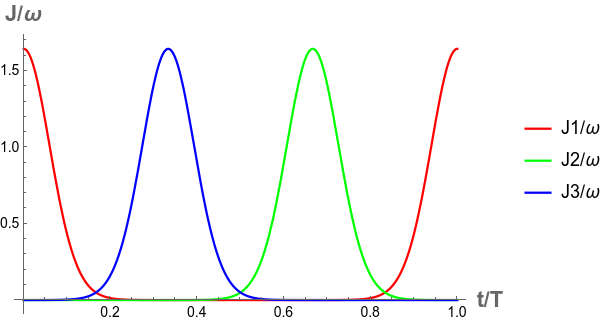} 
\caption{The hoppings $J_l(t)$, $l = 1, 2,3$ from Eq.~\ref{APP:Eqn:hopping_HC} with parameters $A = 0.0015 \omega$, $B = 7$, $C = 0$ as functions of time $t$ for one driving period $T = \frac{2\pi}{\omega}$. } \label{App:Fig:HC_model_hoppings}
\end{figure}

The Floquet model realized in \cite{Anomalous_Honeycomb_APP} can be effectively described by a two band model on a hexagonal lattice with hoppings 
\begin{align}
J_l (t) = A  e^{B \cos(\omega t + (l-1) \phi)} + C, \; \phi  = \frac{2 \pi}{3}, \label{APP:Eqn:hopping_HC}
\end{align}
along the different NN links as sketched in Fig. ~\ref{App:Fig:HC_model}, see also \cite{Anomalous_Honeycomb_SOM_APP}. To guarantee that only hopping along one link is active at a time, the value of $B$ must be taken quite large. For the parameters $A = 0.0015 \omega $, $B = 7$, $C = 0$ chosen in the main text, this is fulfilled to good approximation, but not perfectly, which leads to the deviation of the grey dashed line from one in Fig. 3b of the main text. Fig.~\ref{App:Fig:HC_model_hoppings} shows a plot of the time dependence of $J_l(t)$ for these parameters.

\section{Resonant driving} 
In a Floquet system with nearest-neighbour (NN) hoppings, resonant driving amounts to the conditions that ($i$) only hopping along one NN link is possible at a time and ($ii$) 100~\% intensity transfer is achieved during link activation time. Then, an excitation at some initial site can be pumped along a closed loop perpetually. If ($i$) is fulfilled, ($ii$) is only a matter of scaling the hopping amplitude along the link by a constant prefactor. If two neighbouring sites are coupled by a real time-dependent hopping $J(t)$ that is only non-zero between $t_\mathrm{on}$ and $t_\mathrm{off}$, they are described by the Hamiltonian $H(t) = J(t) \sigma_x$ while $t \in [t_\mathrm{on}, t_\mathrm{off}]$, where $\sigma_x$ is the standard Pauli matrix. Since the Hamiltonian commutes with itself at all times, the time evolution operator can be written as $U(t_\mathrm{off}, t_\mathrm{on}) = \cos(\theta) \sigma_0 - i \sin(\theta) \sigma_x$, where $\Theta = \int_{t_\mathrm{on}}^{t_\mathrm{off}} J(t) \mathrm d t $. The condition for resonant driving is thus

\begin{align}
\int_{t_\mathrm{on}}^{t_\mathrm{off}}  J(t) \mathrm d t = \frac{\pi}{2} + n * \pi \; n \in \mathbb{N},
\end{align}
which can always be met by scaling $J(t)$ with a constant prefactor or by tuning the driving period.

\section{Varying the non-linearity} 
{
Here, we provide additional data to demonstrate the genuinely non-linear nature of the scattering effect. To this end, we let the two pulses collide while varying the non-linearity $\gamma$ and keeping all other parameters fixed.

Fig.~\ref{App:Fig:Phot_flux_vary_NL} shows the result for the photonic square lattice model. Clearly, the dependence on the phase $\Delta$ vanishes for $\gamma \to 0$ and for $\gamma = 0$, the curve is ideally flat.

\begin{figure}[h!]	 
\includegraphics[ width=0.95\linewidth]{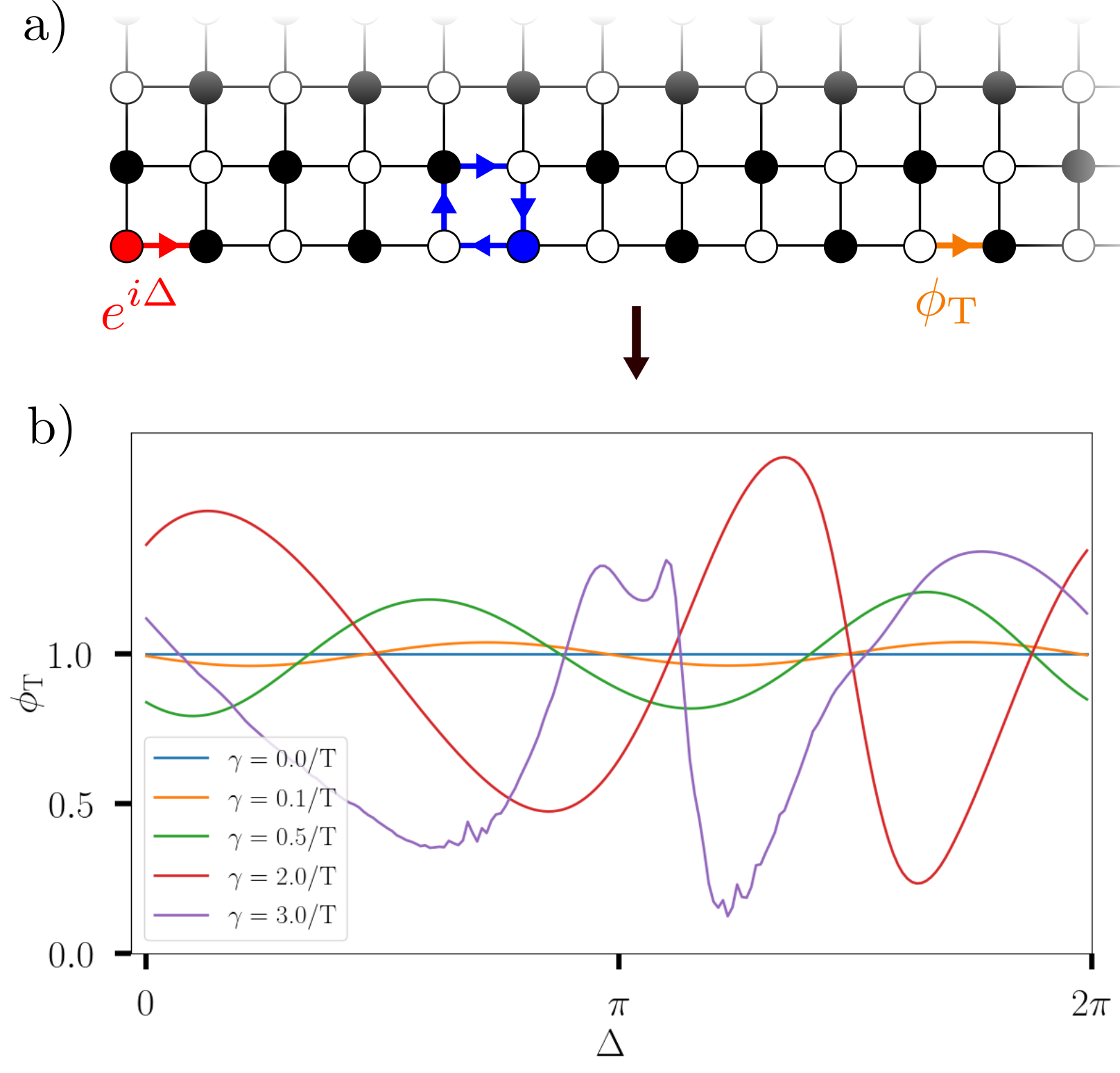} 
\caption{(a) Schematic of square lattice model from the main text with sublattices $A$ and $B$ indicated by white and black circles, respectively. System size is $16 \times 16$ sites and the hopping strength is $c = \frac{2 \pi}{T}$. Initial excitations are located on the lower left edge, with chiral edge state amplitude $\mathrm  e^{i \Delta}$ (red) and the circulating bulk state amplitude $1$ (blue). We measure the transmitted probability flux $\phi_\mathrm{T} = \int_0^{10T} \phi_{\bm j, \bm \delta}(t) \mathrm d t$ as defined in the main text through the link marked in orange. (b) $\phi_\mathrm{T}$ as a function of the relative phase $\Delta$ between the red and blue pulse for different values of the non-linearity $\gamma$.
 } 
\label{App:Fig:Phot_flux_vary_NL}
\end{figure}

Fig.~\ref{App:Fig:HC_flux_vary_NL} shows the result for the Honeycomb model. Here, the dependence on the phase $\Delta$ becomes very small for $\gamma$ approaching zero, but does not vanish entirely. This is because the system is not tuned perfectly to resonant driving, such that a small portion of the blue circulating wave packet is lost over time and propagates along the edge. The non-vanishing $\Delta$-dependence for $\gamma =0$ is due to interference between between this signal and the red wave packet. 

To avoid confusion with the plots shown in Fig. 2 and Fig. 3  of the main text, note that the red and grey dashed lines there are obtained from the red pulse alone, i.e. without starting the circulating blue mode. 

\begin{figure}[h!]	 
\includegraphics[ width=0.95\linewidth]{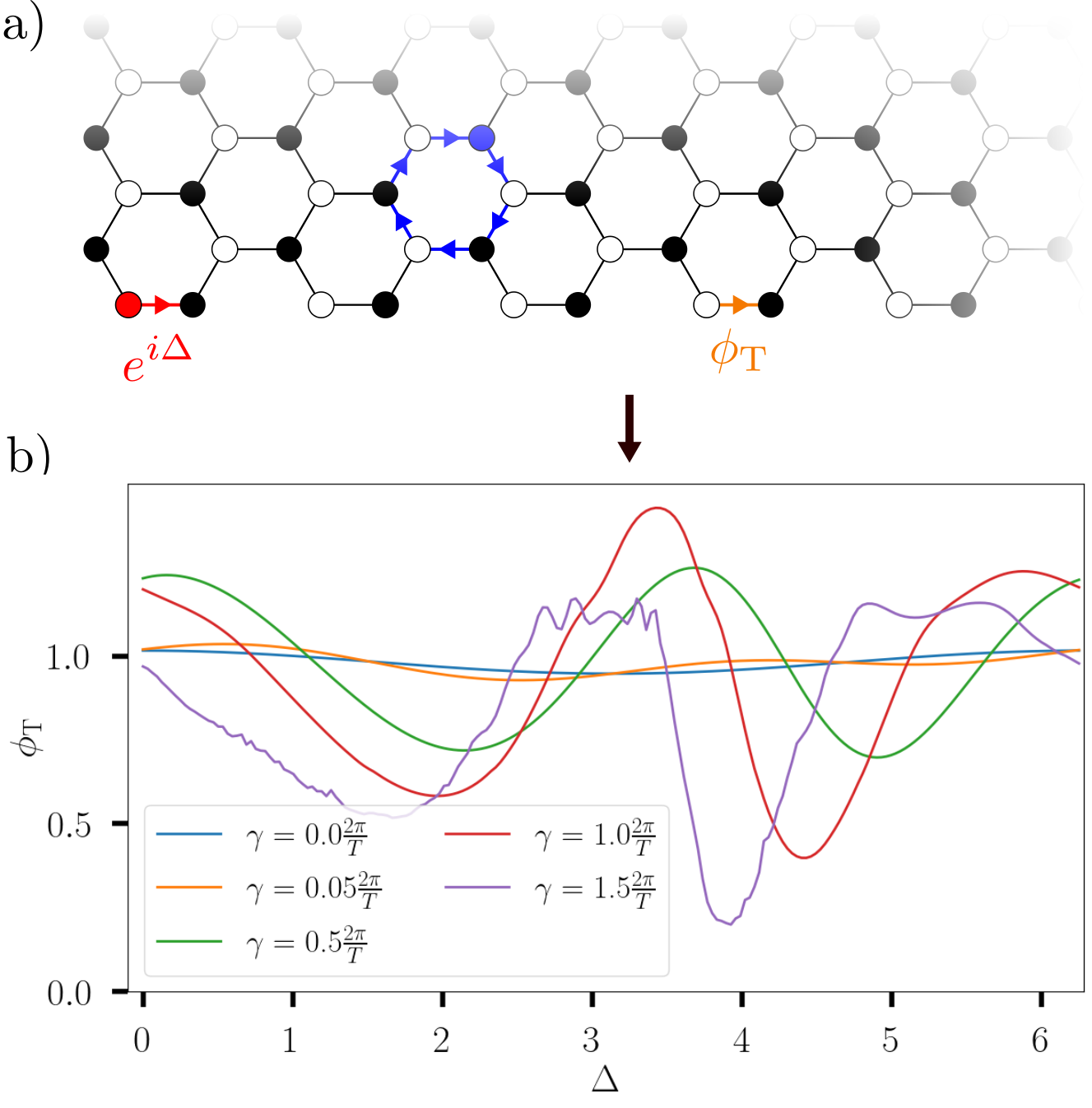} 
\caption{(a) Schematic of Honeycomb lattice model from the main text with sublattices $\alpha$ and $\beta$ indicated by white and black circles, respectively. System size is $10 \times 10$ unit cells and the time dependent hoppings are given by Eq.~(\ref{APP:Eqn:hopping_HC}) with parameters $A = 0.0015\frac{2 \pi}{T}$, $B = 7$, $C = 0$. Initial excitations are located on the lower left edge, with chiral edge state amplitude $\mathrm  e^{i \Delta}$ (red) and the circulating bulk state amplitude $1$ (blue). We measure the transmitted probability flux $\phi_\mathrm{T} = \int_0^{20 T} \phi_{\bm j, \bm \delta}(t) \mathrm d t$ over 20 driving periods $T = \frac{2 \pi}{\omega}$ through the link marked in orange. (b) $\phi_\mathrm{T}$ as a function of the relative phase $\Delta$ between the red and blue pulse for different values of the non-linearity $\gamma$.} \label{App:Fig:HC_flux_vary_NL}
\end{figure}

}

\end{document}